\documentclass[a4paper,english,eqsecnum,a4paper,twocolumn,showpacs,preprintnumbers,amsmath,amssymb,prb]{revtex4}
\usepackage[latin1]{inputenc}
\usepackage{float}
\usepackage{graphicx}
\usepackage{amssymb}

\makeatletter

\usepackage{float}

\makeatletter

\usepackage{float}

\makeatletter

\usepackage{dcolumn}

\usepackage{bm}

\makeatother

\makeatother

\usepackage{babel}
\makeatother
\begin{document}

\title{Elementary excitations probed by
$L$-edge resonant inelastic x-ray scattering
in systems with weak and intermediate electron correlations}

\author{Jun-ichi Igarashi$^{1}$ and Tatsuya Nagao$^{2}$}

\affiliation{
 $^{1}$Faculty of Science, Ibaraki University, Mito, Ibaraki 310-8512,
Japan\\
$^{2}$Faculty of Engineering, Gunma University, Kiryu, Gunma 376-8515,
Japan
}

\date{\today}

\begin{abstract}
We develop a formalism to calculate the 
$L$-edge resonant x-ray scattering 
(RIXS) spectra from transition-metal compounds. Using a multi-orbital 
tight-binding model, we derive useful formulas to calculate the spectra
by collecting up the ladder diagrams on the basis of the Keldysh scheme,
without relying on the fast collision approximation.
They are feasible in the weak and intermediate coupling regimes of itinerant
electron systems, where the charge and the magnetic excitations are mixed 
together. We examine the difference between the present formulas and
those of the fast collision approximation.
Finally, to demonstrate how our theory works, we employ the formulas to
study the RIXS spectra on 
a simple model, the single-orbital Hubbard model 
on the square lattice at half-filling.
The intensities originated from the magnon and from the continuous
states are obtained on the equal footing in the antiferromagnetic ground state.

\end{abstract}


\maketitle

\section{\label{sect.1}Introduction}

Elementary excitations in solids are fundamental to describe physical properties
such as the response to external perturbations and temperature dependence.
Among several probes, one of the most known ones is inelastic neutron 
scattering, by which we could obtain the energy-momentum relations of magnetic 
excitations. Another probe is the optical method, by which we could observe 
charge excitations with their momenta nearly zero.
Recently, taking advantage of strong synchrotron
sources, resonant inelastic x-ray scattering (RIXS) has become
a powerful tool to probe both charge and magnetic excitations in solids.
\cite{Ament11}
Both the $K$- and $L$-edge resonances are utilized in 
transition-metal compounds, where the $K (L)$-edge RIXS is described
as a second-order dipole allowed process that the $1s$ ($2p$) 
core electron is prompted to an empty $4p$ ($3d$) state by 
absorbing photon, then the $4p$ ($3d$) electron is combined 
with the core hole by emitting photon. 
In the end, charge and/or magnetic excitations are left
with energy and momentum transferred from photon.

The $K$-edge resonances\cite{Kao96,Hill98,Hasan00,Kim02,Inami03,Kim04-1,
Suga05,Ishii2007}
are more useful than the $L$-edge 
ones\cite{Ghiringhelli04,Ulrich2008}
in order to observe momentum dependence of the RIXS intensities, 
because the corresponding x-rays have
wavelengths of the same order of lattice spacing.
In undoped cuprates, several spectral peaks have been observed
around $\omega\sim 2-8$ eV.\cite{Hasan00,Kim02,Suga05}
They exhibit characteristic momentum dependence and seem to be
brought about by the charge excitations.
Among several attempts to explain the spectra,
\cite{Tsutsui99,Okada06,Markiewicz06,vdBrink06,Ament07,Vernay2008} 
Nomura and Igarashi (NI)\cite{Nomura04,Nomura05,Igarashi06} 
have developed a formalism on the basis of the Keldysh 
scheme,\cite{Keldysh65} in which the spectra are described 
in terms of the density-density correlation function. 
NI have calculated the spectra on the d-p model 
by collecting up the ladder diagrams, 
and succeeded in semi-quantitatively 
explaining the RIXS spectra for La$_2$CuO$_4$.\cite{Nomura04,Nomura05,Igarashi06} 
This approach has been extended to multi-orbital tight-binding 
models, 
providing quantitative explanations to the spectra for
La$_2$CuO$_4$,\cite{Takahashi08} NiO,\cite{Takahashi07} LaMnO$_3$,
\cite{Semba08} and La$_2$NiO$_4$.\cite{Nomura12}
In addition to these spectra due to the charge excitations,
the RIXS intensities originated from the magnetic excitations 
have been observed at the Cu $K$-edge in La$_2$CuO$_4$ around
400 meV.\cite{Hill08,Braicovich09,Ellis10}
The NI formula has been adapted to treat the process of 
exciting two magnons\cite{Nagao07} on the basis of the 
mechanism that the exchange interaction 
is modified by the core-hole 
potential.\cite{vdBrink07,Forte08}
The result has provided a quantitative explanation by properly 
taking account of the magnon-magnon interaction.

The $L$-edge resonances probe directly the $d$ states, 
although the accessible wave vector is more limited than the 
$K$-edge ones. 
Recent instrumental improvement of the energy resolution
has made it possible to distinguish the magnetic excitations
from the entire spectral peaks at the Cu $L$-edge in undoped
cuprates.
Their energy profile shows asymmetric shape,\cite{Braicovich09,Braicovich10,Guarise10}
which indicates that they are originated 
from not only one magnon but also two magnons.
Although the spectra have been analyzed within the 
fast-collision approximation,\cite{Ament07,Ament09,Haverkort10}
it could explain only the one-magnon contribution.
The present authors have developed a `projection' method 
to go beyond the
fast collision approximation,\cite{Igarashi12-1,Igarashi12-2,Nagao12} 
and have explained quantitatively the spectral shape.
These analyses are based on the localized spin model
and utilize the $1/S$
expansion\cite{Igarashi92-1,Igarashi92-2} to include the 
magnon-magnon interaction.

Recently, magnetic excitations have been observed even in the 
superconducting phase in doped cuprates;\cite{Tacon11}
a low-energy peak is found, which is assigned to be 
brought about by paramagnons. 
This experiment urges us to develop a formalism
describing the magnetic excitations in itinerant electron systems,
since the above analyses in undoped cuprates are based on 
the localized spin models.
Another motivation to develop the formalism for itinerant 
electron systems comes from the RIXS experiment at the 
Ir $L$-edge in Sr$_2$IrO$_4$,\cite{Kim12} where
a low-energy peak as well as continuous spectra are observed.
Although the spectrum has been analyzed by using a model with 
localized spin and orbitals,\cite{Ament11-1} 
it may make sense to analyze the spectra
from the point of view of itinerant electron systems.

In this paper, bearing future analyses of such spectra in mind, 
we develop a theory
to calculate the RIXS spectra for itinerant electron systems on 
the basis of the Keldysh scheme
without relying on the fast collision approximation. 
Then, complication arises from the fact that the creation and 
annihilation of $d$ electron take place at different times, 
which seems different from $K$-edge 
RIXS.\cite{Nomura04,Nomura05,Igarashi06} 
We describe the electronic structure within the Hartree-Fock approximation 
(HFA) by introducing a multi-orbital tight-binding model.
We obtain the formulas to calculate the RIXS spectra by collecting up 
the ladder diagrams, which enables us to treat both the bound 
state and the continuous states on the equal footing 
in the magnetic and/or orbital ordering ground state.
Although analysis based 
on the fast collision approximation 
has been carried out for itinerant electron systems such as
iron arsenide superconductors,\cite{Kaneshita11}
it seems, however, unclear to us whether the assumption made is valid
in the situations where several orbitals with different energies 
should be treated simultaneously.
We examine the difference between the present formulas and those 
of the fast collision approximation.
Finally, to demonstrate how the present theory works, 
we evaluate the RIXS spectra
on a simple model, the single-orbital Hubbard model  
on the square lattice at half-filling, 
where the magnon exists as a bound state 
in the antiferromagnetic ground state. 
Note that the RIXS spectra for the same model in the one dimension have
been studied by the exact diagonalization method.\cite{Kourtis2012}

The present paper is organized as follows. In Sec. \ref{sect.2},
we introduce a multi-orbital tight-binding model, and study the electronic 
structure within the HFA. In Sec. \ref{sect.3}, we develop a formalism 
for the L-edge RIXS spectra within the ladder approximation 
by employing the Keldysh formalism. 
In Sec. \ref{sect.4}, we calculate the spectra
on the single-band Hubbard model at half-filling on the square lattice. 
Section \ref{sect.5} is devoted to the concluding remarks. 
In Appendix \ref{appen.A}, we derive Eq.~(\ref{eq.L}) 
in the time representation, and in Appendix \ref{appen.B},
we discuss the fluctuation-dissipation theorem 
in the present context. 

\section{\label{sect.2}Multi-orbital tight-binding model 
and the Hartree-Fock approximation}

We describe the electronic structures of transition-metal compounds 
by using a multi-orbital tight-binding model defined as
\begin{equation}
H_{mat} =  H_{0}+H_{I},
\end{equation}
with
\begin{eqnarray}
H_{0} & = & \sum_{\left\langle i,i' \right\rangle }
\sum_{n,n'\sigma}\left(t_{in,i'n'}a_{in\sigma}^{\dagger}a_{i'n'\sigma}+H.c.
\right),\\
H_{I} & = & 
 \frac{1}{2} \sum_{i} 
 \sum_{\nu_1\nu_2\nu_3\nu_4}g(\nu_1\nu_2;\nu_3\nu_4)
             a_{i\nu_1}^{\dagger} a_{i\nu_2}^{\dagger}
             a_{i\nu_4} a_{i\nu_3}.
\label{eq.HI}
\end{eqnarray}
The $H_{0}$ represents the kinetic energy with transfer integral 
$t_{in,i'n'}$, where $a_{in\sigma}$ ($a_{in\sigma}^{\dagger}$) 
denotes the annihilation (creation) operator of 
an electron with orbital $n$ and spin $\sigma$ at the transition-metal site 
$i$. The $H_{I}$ represents the on-site Coulomb interaction between electrons, which may be described with the use of Slater integrals.
The label $(n,\sigma)$ specifying the $d$ state is abbreviated 
as $\nu$.

We consider the bipartite lattice in order to take account of the possible 
antiferromagnetic and/or antiferro-orbital order, 
and introduce the Fourier transform of the annihilation operators
in the reduced first Brillouin zone (BZ): 
\begin{equation}
 a_{\lambda\nu }({\bf k}) = \sqrt{\frac{2}{N}}\sum_{i}a_{i\nu}
                             \exp(-i{\bf k}\cdot{\bf r}_i) ,
\label{eq.Fourier}
\end{equation}
with $i$ running over the A and B sublattices for $\lambda=1$ and 
$\lambda=2$, respectively.
Using the abbreviation $\xi=(\lambda,\nu)$, 
$H_0$ may be expressed as
\begin{equation}
 H_{0} = \sum_{{\bf k}\xi\xi'} a_{\xi}^{\dagger}({\bf k})
          \left[\hat{H}_{0}({\bf k})\right]_{\xi,\xi'}
          a_{\xi'}({\bf k}),
\end{equation}
where $\hat{H}_{0}({\bf k})$ includes the Fourier transform 
of $t_{in,i'n'}$,
whose explicit form is omitted here.

Now let us introduce a single-particle Green's function in 
a matrix form 
\begin{equation}
\left[\hat{G}({\bf k},\omega)\right]_{\xi,\xi'}=-i\int\langle 
 T(a_{\xi}({\bf k},t)a_{\xi'}^{\dagger}({\bf k},0))\rangle
 {\rm e}^{i\omega t}{\rm d}t,
\label{eq.dG}
\end{equation}
where $T$ is the time ordering operator, and $\langle X \rangle$
denotes the ground-state average of operator $X$.
In the HFA, we disregard the fluctuation terms in $H_{I}$, which 
leads to 
\begin{equation}
H_{I}^{HF}=\sum_{i}\sum_{\xi_{1}\xi_{2}\xi_{3}\xi_{4}}
\Gamma^{(0)}(\xi_{1}\xi_{2};\xi_{3}\xi_{4})\langle 
a_{i\xi_{2}}^{\dagger}a_{i\xi_{4}}\rangle a_{i\xi_{1}}^{\dagger}a_{i\xi_{3}},
\end{equation}
where $\Gamma^{(0)}$ is the antisymmetric vertex function defined by 
\begin{equation}
\Gamma^{(0)}(\xi_{1}\xi_{2};\xi_{3}\xi_{4})
=g(\xi_{1}\xi_{2};\xi_{3}\xi_{4})-g(\xi_{1}\xi_{2};\xi_{4}\xi_{3}).
\label{eq.Gamma0}
\end{equation}
Here, $\xi_n=(\lambda_n,\nu_n)$ for $n=1 \sim 4$. 
Equation (\ref{eq.Gamma0}) gives non-zero value only when
$\lambda_1=\lambda_2=\lambda_3=\lambda_4$.

Considering the equation of motion with $H_0+H_{I}^{HF}-\mu\hat{N}$,
we obtain
\begin{equation}
(\omega\hat{I}-\hat{J}({\bf k}))\hat{G}({\bf k},\omega)=\hat{I},
\end{equation}
where $\mu$ is the chemical potential, and $\hat{I}$ is the unit matrix.
The $\hat{J}({\bf k})$ is defined 
through the relation
\begin{equation}
 \left[a_{\xi}({\bf k}), H_0+H_{I}^{HF}-\mu \hat{N}\right] 
 =\sum_{\xi'}\left[\hat{J}({\bf k})\right]_{\xi\xi'}a_{\xi'}({\bf k}).
\end{equation}
We diagonalize $\hat{J}({\bf k})$ by a unitary matrix 
$\hat{U}({\bf k})$:
$[\hat{U}({\bf k})^{-1}\hat{J}({\bf k})\hat{U}({\bf k})]_{j,j'}
=E_{j}({\bf k})\delta_{j,j'}$ where $E_j(\textbf{k})$ represents
eigenvalue of $\hat{J}(\textbf{k})$. 
Then, the Green's function may be written as 
\begin{equation}
\hat{G}({\bf k},\omega)=\hat{U}({\bf k})\hat{D}({\bf k},\omega)
\hat{U}({\bf k})^{-1},
\end{equation}
with 
\begin{equation}
[\hat{D}({\bf k},\omega)]_{j,j'}=\frac{1}{\omega-E_{j}({\bf k})
+i\delta{\rm sgn}[E_{j}({\bf k})]}\delta_{j,j'},
\end{equation}
where ${\rm sgn}[A]$ stands for a sign of quantity $A$ and $\delta$ denotes
a positive convergent factor.
The expectation values of the electron density operator on the ground state, 
which are contained in $\hat{J}({\bf k})$,
are self-consistently determined from
\begin{equation}
\langle a_{\xi}^{\dagger}a_{\xi'}\rangle=
\frac{2}{N}\sum_{{\bf k}}(-i)\int
[\hat{G}({\bf k},\omega)]_{\xi,\xi'}
{\rm e}^{i\omega0^{+}}\frac{{\rm d}\omega}{2\pi}. 
\label{eq.gap}
\end{equation}
We assume that there exists a stable self-consistent solution 
such as the antiferromagnetic and/or the antiferro-orbital order.

\section{\label{sect.3}Formulation of RIXS spectra}

\subsection{Dipole transition}

For the interaction between photon and matter, we consider the dipole
transition at the $L$ edge, where the $2p$-core electron is excited
to the $d$ states by absorbing photon (and the reverse process).
The $2p$-core states are split into the multiplet of angular 
momentum $j=3/2$ and $1/2$ due to the spin-orbit coupling. 
Introducing the annihilation operator $p_{i;jm}$ 
of the core $2p$ electron with the magnetic 
quantum number $m$, we write the Hamiltonian for the core electron as
\begin{equation}
H_{2p} = \sum_{jm}\epsilon_{2p}(j)\sum_{\lambda,{\bf k}}
p_{\lambda,jm}^{\dagger}({\bf k}) p_{\lambda,jm}({\bf k}),
\end{equation}
where $p_{\lambda,jm}({\bf k})$ is the Fourier transform of $p_{i;jm}$  
defined similarly by Eq.~(\ref{eq.Fourier}).
The dipole transition may be described by the interaction 
\begin{eqnarray}
H_{x}&=&\sum_{\lambda,n\sigma,jm,\alpha}w(n\sigma;jm;\alpha)\nonumber \\
&\times& \sum_{\bf{k},\bf{q}}
a_{\lambda n\sigma}^{\dagger}({\bf k+q})p_{\lambda,jm}({\bf k})
c_{\alpha}({\bf q})+{\rm H.c.},
\end{eqnarray}
where $c_{\alpha}({\bf q})$ is the annihilation
operator of photon with momentum ${\bf q}$ and polarization $\alpha$.
The $w(n\sigma;jm;\alpha)$ represents the matrix element of 
the $2p\to d$ dipole transition. It is defined at each lattice site, 
since the $2p$ states are well localized.
The Hamiltonian of photon may be expressed as
\begin{equation}
 H_{ph} = \sum_{{\bf q}\alpha}\omega_{\bf q}c_{\alpha}^{\dagger}({\bf q})
 c_{\alpha}({\bf q}).
\end{equation}

\subsection{Keldysh formalism}

\begin{figure}
\includegraphics[width=8.0cm]{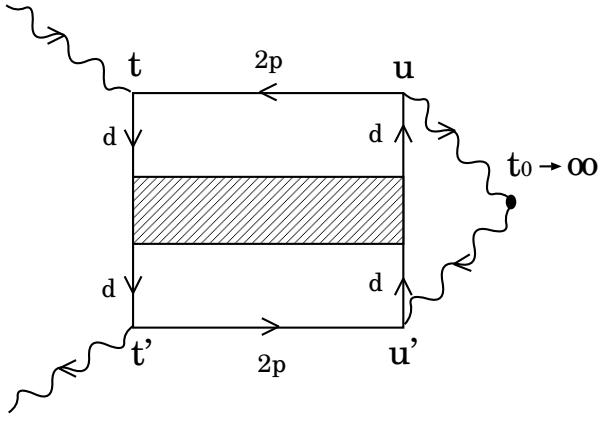}%
\caption{\label{fig.diagram1}
Schematic representation of the RIXS process.
The wavy lines indicate the incident and the scattered photons.
The solid lines with ``2p" and ``d" represent
the Green's functions of
the $2p$-core electron and the $d$ electron, respectively. 
The shaded square represents a general four-point vertex.
}
\end{figure}

We use a Keldysh-Green's function formula to derive a 
general expression 
for the $L$-edge RIXS spectra by extending 
the formalism for the $K$-edge
RIXS spectra given by NI.\cite{Nomura05,Igarashi06}
We set an initial state such that 
one photon exists with momentum ${\bf q}_i$, frequency $\omega_i$,
and polarization $\alpha_i$ in addition to a material in the ground state, 
which may be expressed as
\begin{equation}
  |\Phi_i\rangle = c_{\alpha_i}^{\dagger}({\bf q}_i) |g\rangle , 
\end{equation}
where $|g\rangle$ is the ground state of the matter with no photon.
Let $H\equiv H_{ph}+H_{mat}+H_{2p}$ be the unperturbed Hamiltonian 
of the system and $H_x$ be the perturbation. The $S$ matrix is generally
given by
\begin{equation}
 U(t,-\infty) = T\exp\left\{-i\int_{-\infty}^t H_x(t'){\rm d}t'\right\},
\end{equation}
with $H_x(t)=\exp(iHt)H_x\exp(-iHt)$.
The probability of finding a photon with momentum ${\bf q}_f$, 
frequency $\omega_f$, and polarization 
$\alpha_f$ at time $t_0$ is given by
\begin{equation}
 P_{{\bf q}_f\alpha_f;{\bf q}_i\alpha_i}(t_0)
  =\langle \Phi_i| U(-\infty,t_0)c_{{\bf q}_f\alpha_f}^\dagger
              c_{{\bf q}_f\alpha_f}U(t_0,-\infty) |\Phi_i\rangle .
\label{eq.prob}
\end{equation}
Time $t_0$ is set $t_0\rightarrow \infty$ at the end, since the scattered
photon is to be observed far after the scattering event takes place.
We expand the S-matrix to second order in $H_x$, 
\begin{eqnarray}
 U(t,-\infty) &=& 1 + (-i)\int_{-\infty}^t H_x(t'){\rm d}t' \nonumber\\
              &+& \frac{(-i)^2}{2}\int_{-\infty}^t \int_{-\infty}^t
	        T(H_x(t')H_x(t'')){\rm d}t'{\rm d}t''.
\nonumber \\
\end{eqnarray}
Inserting this into Eq.~(\ref{eq.prob}), and factoring out the dependence 
on the photon frequencies, we obtain
\begin{eqnarray}
 P_{q_f\alpha_f;q_i\alpha_i}(t_0)
  &=& \int_{-\infty}^{t_0}{\rm d}u\int_{-\infty}^{u}{\rm d}t
    \int_{-\infty}^{t_0}{\rm d}u'\int_{-\infty}^{u'}{\rm d}t'\nonumber\\
&\times&
     S(t,u;t',u'){\rm e}^{i\omega_i(t'-t)}{\rm e}^{-i\omega_f(u'-u)},
\end{eqnarray} 
where $S(t,u;t',u')$ is the contribution of all the electron lines and
electron-hole vertices as illustrated in Fig.~\ref{fig.diagram1}.
The wavy lines indicate photons, which carry energy and momentum.
The solid lines with ``2p" and ``d" represent 
the Green's function of the $2p$-core and $d$ electrons, respectively.
The upper and lower halves of the graph correspond to the so called
``outward" and ``backward" time legs, respectively.
The transition probability per unit time could be obtained from
this expression by letting $t_{0}\to\infty$ and 
fixing one time, for instance, 
$u=0$:\cite{Nozieres74}
\begin{eqnarray}
 W(q_f\alpha_f;q_i\alpha_i)
  &=& \int_{-\infty}^{0}{\rm d}t
    \int_{-\infty}^{\infty}{\rm d}u'\int_{-\infty}^{u'}{\rm d}t' \nonumber \\
&\times& 
     S(t,0;t',u'){\rm e}^{i\omega_i(t'-t)}{\rm e}^{-i\omega_fu'}.
\label{eq.general}
\end{eqnarray} 

\subsection{Amplitudes creating an electron-hole pair excitation}

Now we evaluate Eq.~(\ref{eq.general}) for the $L_3$ edge
on the basis of the Keldysh formalism, where four kinds of
the Green's functions $G^{--}$, $G^{-+}$, $G^{+-}$, and 
$G^{++}$ are utilized;
the superscripts $+$ and $-$ of the Green's functions indicate the backward 
and outward time legs, respectively.\cite{Landau} 
For example, $[\hat{G}^{--}({\bf k},\omega)]_{\xi,\xi'}$ is the same 
as defined in Eq. (\ref{eq.dG}), and the Green's function 
$G_{2p}^{--}({\bf k},\omega)$ of the 2p-core electron 
may be given by
\begin{equation}
 G_{2p}^{--}({\bf k},\omega)= \frac{1}{\omega-\epsilon_{2p}(3/2)-i\Gamma_c},
\end{equation}
where $\Gamma_c$ represents the life-time broadening width of the core hole.
The expansion technique with respect to the Coulomb interaction
is well explained in Ref.~\onlinecite{Landau}. 
We use the same notation as given in 
Ref.~\onlinecite{Landau}. The unperturbed Green's function of the 
$d$ electron is given by the HFA.

\begin{figure}
\includegraphics[width=8.0cm]{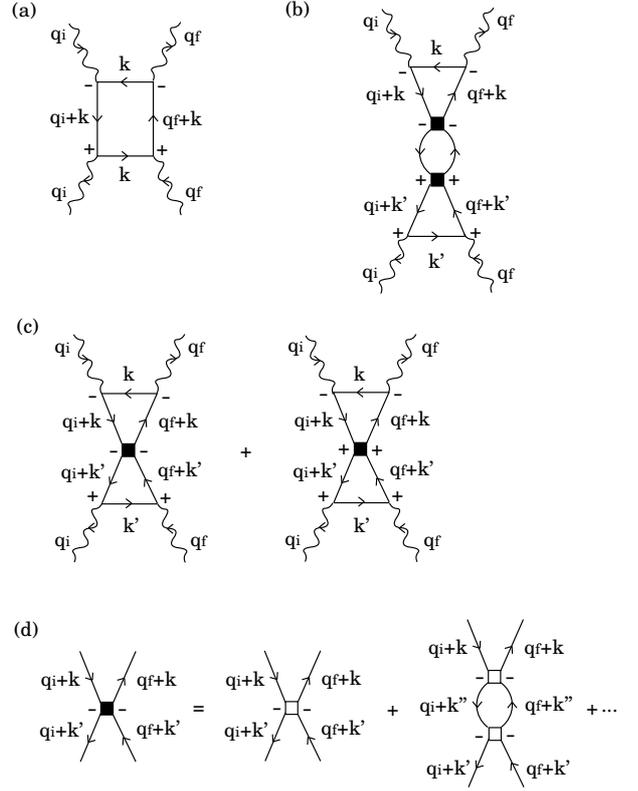} 
\caption{\label{fig.diagram2} 
Diagrams within the ladder approximation.
Panels (a) amd (b) are the diagrams of the 
lowest-order and the higher-order, respectively, 
in the Coulomb interaction. 
Panel(c) is the diagram where the renormalized
vertex works only at the outward time leg and 
at the backward time leg, respectively.
Panel (d) represents the renormalized vertex which is the sum of
the ladder diagrams.
The horizontal and vertical solid lines represent 
the Green's functions of the 
2p-core and $d$ electrons, respectively.
The filled squares represent the renormalized vertex.
The symbols $-$ and $+$ indicate the outward and backward time legs,
respectively.
}
\end{figure}

We first consider the lowest-order diagram shown in Fig.~\ref{fig.diagram2}(a).
Carrying out the integration with respect to 
the frequency of the 2p-Green's function, 
we obtain the RIXS intensity at the $L_3$ edge:
\begin{eqnarray}
 &&W_a(\omega_i,q;\alpha_i,\alpha_f) \nonumber \\
&=& 2\pi
  \sum_{\xi\xi'\xi_{1}\xi'_{1}}
  \frac{2}{N}\sum_{{\bf k}}\sum_{j\ell}
  \left|R(\omega_i,E_j({\bf k+q}))\right|^2 
  M^*(\xi_1 \xi'_1;\alpha_i,\alpha_f)
    \nonumber \\
 &\times&
  U_{\xi_1 j}({\bf k+q})
  U^*_{\xi j}({\bf k+q}) U_{\xi' \ell}({\bf k})U^*_{\xi'_1 \ell}({\bf k}) 
\left[1-n_j({\bf k+q})\right]
  \nonumber\\
 &\times& n_{\ell}({\bf k})
 \delta(\omega-E_j({\bf k+q})+E_{\ell}({\bf k}))
  M(\xi\xi';\alpha_i,\alpha_f) ,
\label{eq.spec.a}
\end{eqnarray}
with
\begin{eqnarray}
   R(\omega_i,E) &\equiv& 
  \frac{1}{\omega_i-E+\epsilon_{2p}(3/2)+i\Gamma_c}, \\
  M(\xi\xi';\alpha_i,\alpha_f)
   &=& \sum_{m} w(\xi;3/2,m;\alpha_i)w^*(\xi';3/2,m;\alpha_f), \nonumber \\
\label{eq.M}
\end{eqnarray}
where $q_i\equiv (\omega_i,{\bf q}_i)$, 
$q_f\equiv (\omega_f,{\bf q}_f)$, and
$q\equiv (\omega,{\bf q})=q_i-q_f$. Here ${\bf k+q}_f$ is replaced 
by ${\bf k}$, which runs over the reduced first BZ.
The $n_{\ell}({\bf k})$ denotes the occupation number of the eigenstate 
with energy $E_{\ell}({\bf k})$, 
The $\delta$-function indicates the conservation of energy for the continuum
states of an electron-hole pair excitation.

%

Second, we consider the diagram shown in Fig.~\ref{fig.diagram2}(b).
The solid square represents the renormalized vertex function, which 
we evaluate by collecting up the ladder diagrams.
Then, carrying out the integration with respect to the frequency 
of the 2p Green's function, 
we obtain the contribution to the RIXS intensity as
\begin{eqnarray}
&& W_b(\omega_i,q;\alpha_i,\alpha_f) \nonumber \\
&=& 
 \hat{M}^{\dagger}(\alpha_i,\alpha_f)\hat{L}^{\dagger}(\omega_i,q)
  \hat{\Gamma}
 \left[\hat{I}+\hat{F}^{\dagger}(q)\hat{\Gamma}\right]^{-1}
 \hat{\Pi}^{(0)}(q) \nonumber\\
 &\times&\left[\hat{I}+\hat{\Gamma}\hat{F}(q)\right]^{-1}
 \hat{\Gamma}\hat{L}(\omega_i,q)\hat{M}(\alpha_i,\alpha_f),
\label{eq.spec.b}
\end{eqnarray} 
where each factors are given in the matrix form
represented with the base states $\xi\xi'$. 
The $\hat{M}$ is the matrix form of Eq.~(\ref{eq.M}):
\begin{equation}
 \left[\hat{M}(\alpha_i,\alpha_f)\right]_{\xi\xi'} = 
 M(\xi\xi';\alpha_i,\alpha_f).
\end{equation}
\begin{widetext}
The $\hat{L}$ corresponds to the diagram above the renormalized vertex
on Fig.~\ref{fig.diagram2}(b):
\begin{eqnarray}
&& \left[\hat{L}(\omega_i,q)\right]_{\xi_1\xi'_{1};\xi\xi'} 
=
 \frac{2}{N}\sum_{{\bf k}}\sum_{j,\ell}
     U_{\xi_1 j}({\bf k+q})U^*_{\xi j}({\bf k+q}) 
     U_{\xi' \ell}({\bf k})U^*_{\xi'_1 \ell}({\bf k}) 
      R(\omega_i,E_j({\bf k+q}))
  \nonumber \\
 &\times& 
 \Biggl\{ \frac{1-n_{\ell}({\bf k})}
 {\omega_f-E_{\ell}({\bf k})+\epsilon_{2p}(3/2)+i\Gamma_c}
-\Biggl[
 \frac{[1-n_{j}({\bf k+q})]n_{\ell}({\bf k})}
 {\omega-E_{j}({\bf k+q})+E_{\ell}({\bf k})+i\delta}
  -\frac{n_{j}({\bf k+q})[1-n_{\ell}({\bf k})]}
 {\omega-E_{j}({\bf k+q})+E_{\ell}({\bf k})-i\delta}\Biggr]\Biggr\}. 
\label{eq.L}
\end{eqnarray}
We get a complicated form for $\hat{L}(\omega_i;q)$, 
since the creation and annihilation of $d$ electron take place at different
times. For understanding the origin of each term, it may be helpful to carry 
out the integration in the time representation (See Appendix \ref{appen.A}).
Vertex $\hat{\Gamma}$ is defined by 
\begin{equation}
[\hat{\Gamma}]_{\xi_{2}\xi'_{2};\xi_{1}\xi'_{1}}=
\Gamma^{(0)}(\xi_{2}\xi'_{1};\xi_{1}\xi'_{2}).
\end{equation}
The $\hat{F}(q)$ is given by
\begin{eqnarray}
[\hat{F}(q)]_{\xi_2\xi'_{2};\xi_1\xi'_{1}} & = & 
-i\frac{2}{N}\sum_{{\bf k}}\int\frac{{\rm d}k_{0}}{2\pi}
[\hat{G}({\bf k+q},k_{0}+\omega)]_{\xi_{2}\xi_{1}}
[\hat{G}({\bf k},k_{0})]_{\xi'_{1}\xi'_{2}}
\nonumber \\
 & = & \frac{2}{N}\sum_{{\bf k}}\sum_{j,\ell}
 U_{\xi_{2}j}({\bf k+q})U_{\xi_{1}j}^{*}({\bf k+q})
 U_{\xi'_{1}\ell}({\bf k})U_{\xi'_{2}\ell}^{*}({\bf k})
\nonumber \\
 & \times & \left[\frac{[1-n_{j}({\bf k+q})]n_{\ell}({\bf k})}
 {\omega-E_{j}({\bf k+q})+E_{\ell}({\bf k})+i\delta}
 -\frac{n_{j}({\bf k+q})[1-n_{\ell}({\bf k})]}
 {\omega-E_{j}({\bf k+q})+E_{\ell}({\bf k})-i\delta}\right],
 \label{eq.green_pair}
\end{eqnarray}
which is a conventional particle-hole propagator.
Note that the factors $[\hat{I}+\hat{\Gamma}\hat{F}(q)]^{-1}\hat{\Gamma}$ 
and $\hat{\Gamma}[\hat{I}+\hat{F}^{\dagger}(q)\hat{\Gamma}]^{-1}$ are the
renormalized vertices within the ladder approximation, which could contain 
a pole at some $q$, as the indication of the bound state.
As shown in the next subsection, the bound state could surely contribute to
the RIXS intensity.
The $\hat{\Pi}^{(0)}$ corresponds to the diagram between two renormalized
vertices on Fig.~\ref{fig.diagram2}(b):
\begin{eqnarray}
 \left[\hat{\Pi}^{(0)}(q)\right]_{\xi_2\xi'_{2};\xi_1\xi'_{1}} 
 & = & 2\pi\frac{2}{N}\sum_{{\bf k}}\sum_{j,\ell}
 U_{\xi_{2}j}({\bf k+q})U_{\xi_{1}j}^{*}({\bf k+q})
 U_{\xi'_{1}\ell}({\bf k})U_{\xi'_{2} \ell}^{*}({\bf k}) \nonumber\\
 &\times& 
 [1-n_{j}({\bf k+q})]n_{\ell}({\bf k})
 \delta(\omega-E_{j}({\bf k+q})+E_{\ell}({\bf k})).
\label{eq.pi0}
\end{eqnarray}

To complete the ladder approximation, we need to include another type of 
diagrams shown in Fig. \ref{fig.diagram2}(c). 
Carrying out the integration with respect to 
the frequency of the 2p Green's function,
we obtain the contribution from the diagrams in Fig.~\ref{fig.diagram2}(c):
\begin{eqnarray}
W_c(\omega_i,q;\alpha_i,\alpha_f) &=& 
  \hat{M}^{\dagger}(\alpha_i,\alpha_f)
 \Bigl\{\hat{N}^{\dagger}(\omega_i,q)
 [\hat{I}+\hat{\Gamma}\hat{F}(q)]^{-1}
\hat{\Gamma}\hat{L}(\omega_i,q)\nonumber\\
 &+& \hat{L}^{\dagger}(\omega_i,q)\hat{\Gamma}
[\hat{I}+\hat{F}^{\dagger}(q)\hat{\Gamma}]^{-1}
\hat{N}(\omega_i,q)
 \Bigr\}\hat{M}(\alpha_i,\alpha_f).
\label{eq.spec.c}
\end{eqnarray}
The $\hat{N}$ corresponds to the diagram above the renormalized vertex
on the right-hand side of Fig.~\ref{fig.diagram2}(c):
\begin{eqnarray}
 \left[\hat{N}(\omega_i,q)\right]_{\xi_1\xi'_{1};\xi\xi'} &=& 2\pi 
 \frac{2}{N}\sum_{{\bf k}}\sum_{j,\ell}
     U_{\xi_1 j}({\bf k+q})U^*_{\xi j}({\bf k+q}) 
     U_{\xi' \ell}({\bf k})U^*_{\xi'_1 \ell}({\bf k}) 
     R(\omega_i,E_j({\bf k+q}))
  \nonumber \\
 &\times&
  [1-n_{j}({\bf k+q})]n_{\ell}({\bf k})
   \delta(\omega-E_j({\bf k+q})+E_{\ell}({\bf k})) .
\label{eq.N}
\end{eqnarray}
Note that the pole in $[\hat{I}+\hat{F}^{\dagger}(q)\hat{\Gamma}]^{-1}$
and $[\hat{I}+\hat{\Gamma}\hat{F}(q)]^{-1}$ in Eq.~(\ref{eq.spec.c})
could not contribute to the RIXS intensity because of 
$\delta(\omega-E_j({\bf k+q})+E_{\ell}({\bf k}))$ in Eq. (\ref{eq.N}).
\end{widetext}

The total RIXS intensity is given by
$W_a(\omega_i,q;\alpha_i,\alpha_f)+W_b(\omega_i,q;\alpha_i,\alpha_f)
+W_c(\omega_i,q;\alpha_i,\alpha_f)$.

\subsection{Correlation function and bound state}

The contribution from the diagram of type (b) 
in Fig. \ref{fig.diagram2} may be rewritten as
\begin{eqnarray}
 W_b(\omega_i,q;\alpha_i,\alpha_f) &=& 
 \hat{M}^{\dagger}(\alpha_i,\alpha_f)\hat{L}^{\dagger}(\omega_i,q)
  \hat{\Gamma}\hat{Y}^{+-}(q)
\nonumber \\
&\times& \hat{\Gamma}\hat{L}(\omega_i,q)\hat{M}(\alpha_i,\alpha_f),
\label{eq.spec.b2}
\end{eqnarray} 
where
\begin{equation}
\left[\hat{Y}^{+-}({\bf q},\omega)\right]_{\xi_1\xi'_{1};\xi\xi'}  =  
\int_{-\infty}^{\infty} 
\langle(\rho_{{\bf q}\xi_1\xi'_1}(t))^{\dagger}
       \rho_{{\bf q}\xi\xi'}(0)\rangle {\rm e}^{i\omega t}{\rm d}t ,
\end{equation}
with 
\begin{equation}
  \rho_{{\bf q}\xi\xi'} = \sqrt{\frac{2}{N}}\sum_{\bf k}
   a_{\xi}^{\dagger}({\bf k+q})a_{\xi'}({\bf k}).
\end{equation}
Function $\hat{Y}^{+-}({\bf q},\omega)$, 
connecting the outward and backward time legs, is nothing but
a density-density correlation function in the equilibrium state. 
Note that the factor $\hat{\Gamma}\hat{L}(\omega_i,q)\hat{M}(\alpha_i,\alpha_f)$
may be regarded as an effective transition amplitude creating an electron-hole 
pair.

In order to prove Eq.~(\ref{eq.spec.b2}), we introduce the time-ordered Green's 
function in the Feynman-Dyson scheme,
\begin{equation}
\left[\hat{Y}^{{\rm T}}(q)\right]_{\xi_1\xi'_{1};\xi\xi'}=-i\int \langle 
T[(\rho_{{\bf q}\xi_1\xi'_{1}}(t))^{\dagger}
\rho_{{\bf q}\xi\xi'}(0)]\rangle
{\rm e}^{i\omega t}{\rm d}t,
\label{eq.green_time}
\end{equation}
where the time-ordering is defined in the outward time leg.
This function is related to the correlation function by using a modified
form of the fluctuation-dissipation theorem for $\omega>0$ 
(see Appendix B):
\begin{equation}
 \left[\hat{Y}^{+-}(q)\right]_{\xi_1\xi'_{1};\xi\xi'}=
 -i\left\{\left[\hat{Y}^{{\rm T}}(q)
 \right]^{*}_{\xi\xi';\xi_1\xi'_{1}}
        -\left[\hat{Y}^{{\rm T}}(q)\right]_{\xi_1\xi'_{1};\xi\xi'}
\right\}.
\label{eq.fdt1}
\end{equation}
Within the ladder approximation in the Feynman-Dyson 
scheme, $\hat{Y}^{T}(q)$ is expressed as
\begin{equation}
\hat{Y}^{{\rm T}}(q)= \hat{F}(q)[\hat{I}+\hat{\Gamma}\hat{F}(q)]^{-1}.
\label{eq.time_ladder}
\end{equation}
Using a relation $1/(\omega-E\pm i\delta)=P\{1/(\omega-E)\}\mp 
i \pi \delta(\omega-E)$ in the last line of Eq.~(\ref{eq.green_pair}),
we express $\hat{F}(q)$ as
\begin{equation}
\hat{F}(q)=\hat{F}_{1}(q)+i\hat{F}_{2}(q),
\end{equation}
where $\hat{F}_{1}(q)$ and $\hat{F}_{2}(q)$ are Hermitian matrices.
Using this Hermitian property and the fact that $\hat{\Gamma}$ is a real 
and symmetric matrix, we find a relation 
\begin{eqnarray}
&&\left[\hat{Y}^{{\rm T}}(q)\right]^{*}_{\xi\xi';\xi_1\xi'_{1}} 
\nonumber \\
&=& 
\left\{[\hat{I}+(\hat{F}_{1}(q)-i\hat{F}_{2}(q))\hat{\Gamma}]^{-1}
(\hat{F}_{1}(q)-i\hat{F}_{2}(q))\right\}_{\xi_1\xi'_{1};\xi\xi'},
\nonumber \\
\end{eqnarray}
and hence
\begin{eqnarray}
&&\left[\hat{Y}^{+-}(q)\right]_{\xi_1\xi'_{1};\xi\xi'} \nonumber \\
&=&
-i\left\{\left[\hat{Y}^{{\rm T}}(q)\right]^{*}_{\xi\xi';\xi_1\xi'_{1}}
-\left[\hat{Y}^{{\rm T}}(q)\right]_{\xi_1\xi'_{1};\xi\xi'}\right\}\nonumber\\
&=&\Bigl\{[\hat{I}+(\hat{F}(q))^{\dagger}\hat{\Gamma}]^{-1}
(-2)\hat{F}_{2}(q)[\hat{I}+\hat{\Gamma}\hat{F}(q)]^{-1}
\Bigr\}_{\xi_1\xi'_{1};\xi\xi'}. \nonumber \\
\label{eq.fdt2}
\end{eqnarray}
Since $-2\hat{F}_{2}(q)$ is equivalent to $\hat{\Pi}^{(0)}(q)$,
we see Eq. (\ref{eq.spec.b2}) is equivalent to Eq.~(\ref{eq.spec.b}).

Now we discuss the bound state. The Green's function given by
Eq.~(\ref{eq.time_ladder}) may have poles for some frequencies
below the energy continuum of an electron-hole pair excitation. 
In the antiferromagetic ground state, for example, 
such bound states are known as ``magnon", which give rise to extra RIXS 
intensities. 
Noting that $\hat{F}_{2}(q)=0$ for $\omega$ below the energy continuum, 
Eq.~(\ref{eq.time_ladder}) is rewritten as 
\begin{equation}
\hat{Y}^{{\rm T}}(q)= \left[\hat{F}_1(q)^{-1}+\hat{\Gamma}\right]^{-1}.
\label{eq.bound1}
\end{equation}
\begin{widetext}
Diagonalizing $\hat{F}_{1}(q)^{-1}+\hat{\Gamma}$ by a unitary matrix,
we assume that one eigenvalue becomes zero at $\omega=\omega_{B}({\bf q})$ 
with the eigenvector $B_{\xi_{2}\xi'_{2}}({\bf q})$. We could expand 
$[\hat{Y}^{{\rm T}}(q)]_{\xi_1\xi'_{1};\xi\xi'}$ around 
$\omega\sim\omega_{B}({\bf q})$ as 
\begin{equation}
\left[\hat{Y}^{{\rm T}}(q)\right]_{\xi_1\xi'_{1};\xi\xi'}=
\frac{[\hat{C}({\bf q})]_{\xi_1\xi'_{1};\xi\xi'}}{\omega-\omega_{B}({\bf q})+i\delta},
\label{eq.bound2}
\end{equation}
with 
\begin{equation}
[\hat{C}({\bf q})]_{\xi_1\xi'_{1};\xi\xi'}=\frac{B_{\xi_1\xi'_{1}}({\bf q})
B_{\xi\xi'}^{*}({\bf q})}
{\sum_{\xi_2\xi'_{2}\xi_3\xi'_{3}}B_{\xi_3\xi'_{3}}^{*}({\bf q})
\frac{\partial}{\partial\omega}[\hat{F}_1({\bf q},\omega_B({\bf q}))^{-1}
]_{\xi_3\xi'_{3};\xi_2\xi'_{2}}B_{\xi_2\xi'_{2}}({\bf q})}.
\label{eq.wt.c}
\end{equation}
\end{widetext}
Substituting Eq.~(\ref{eq.bound2}) into the right hand side 
of Eq.~(\ref{eq.fdt1}), we obtain 
\begin{equation}
 \hat{Y}^{+-}(q)=2\pi
\hat{C}({\bf q})\delta(\omega-\omega_{B}({\bf q})).
\label{eq.y.bound}
\end{equation}
Inserting this result into Eq.~(\ref{eq.spec.b2}),
we have the contribution from the bound state.

\subsection{Fast collision approximation}

The RIXS spectra could be described as the second-order dipole allowed
process. For that process the fast collision approximation 
replaces the intermediate state by a single state.
In the context of the present theory, 
this means that $E_j({\bf k+q})$ is replaced by a certain energy $E_0$
in the denominator of Eq. (\ref{eq.spec.a}), and that factor is decoupled 
from the others. Hence, we have 
\begin{equation}
 W_a(\omega_i,q;\alpha_i,\alpha_f) = 
  \left|R(\omega_i,E_0)\right|^2
  \hat{M}^{\dagger}(\alpha_i,\alpha_f)\hat{\Pi}^{(0)}(q)
  \hat{M}(\alpha_i,\alpha_f).
\end{equation}
In addition, we disregard the first term in Eq.~(\ref{eq.L}), 
and make the same decoupling procedure to the second term.
Hence we have for $\hat{L}(\omega_i,q)$, 
\begin{equation}
 \hat{L}(\omega_i,q) = - R(\omega_i,E_0) \hat{F}(q).
\end{equation}
In the same spirit, we can approximate $\hat{N}(\omega_i,q)$ by
\begin{equation}
 \hat{N}(\omega_i,q) = R(\omega_i,E_0) \hat{\Pi}^{(0)}(q).
\end{equation}
With these approximate expressions, we have the total RIXS intensity as
\begin{eqnarray}
 & & W_a(\omega_i,q;\alpha_i,\alpha_f)+W_b(\omega_i,q;\alpha_i,\alpha_f)+
 W_c(\omega_i,q;\alpha_i,\alpha_f)\nonumber\\
 &=& 
\left|R(\omega_i,E_0)\right|^{2}
 \hat{M}^{\dagger}(\alpha_i,\alpha_f)
 \left[\hat{I}+\hat{F}^{\dagger}(q) \right. \nonumber \\
&\times& \left. 
\hat{\Gamma}\right]^{-1}\hat{\Pi}^{(0)}(q)
 \left[\hat{I}+\hat{\Gamma}\hat{F}(q)\right]^{-1}\hat{M}(\alpha_i,\alpha_f)
\nonumber\\
 &=& \left|R(\omega_i,E_0)\right|^{2}
 \hat{M}^{\dagger}(\alpha_i,\alpha_f)Y^{+-}(q)
          \hat{M}(\alpha_i,\alpha_f) .
\label{eq.int.fast}
\end{eqnarray}
In the situation that $E_j({\bf k+q})$ varies considerably depending on $j$ 
and ${\bf k+q}$, the procedure made above seems difficult to be justified.

\section{\label{sect.4}Application to the single-orbital Hubbard model
on the square lattice}

We have obtained the formulas of the $L$-edge RIXS spectra for multi-orbital 
models. Since they are rather complicated, we demonstrate in the following 
how the formulas work through the application
to a simple model, a single-orbital Hubbard model with nearest-neighbor hopping
on the square lattice at half-filling. 
The Hamiltonian is defined by
\begin{equation}
 H_{\rm mat}=\sum_{\sigma}\sum_{i,j}t_{ij}a_{i\sigma}^{\dagger}a_{j\sigma}
  + U\sum_{i} a_{i\uparrow}^{\dagger}a_{i\downarrow}^{\dagger}
              a_{i\downarrow}a_{i\uparrow} .
\end{equation}

\subsection{HFA to the single-particle states}

The antiferromagetic ordering is known to be realized in the ground state.
Specifying the sublattices by A and B, we put the base states 
$\xi$'s  in order ($A,\uparrow$), ($B,\uparrow$), ($A,\downarrow$), 
and ($B,\downarrow$).
Then $\hat{J}({\bf k})$ is represented as
\begin{equation}
 \hat{J}({\bf k}) =
   \left( \begin{array}{cccc}
      -Um & \epsilon_{\bf k} & 0 & 0 \\
      \epsilon_{\bf k} & Um  & 0 & 0 \\
       0 & 0 & Um & \epsilon_{\bf k} \\
       0 & 0 & \epsilon_{\bf k} & -Um 
          \end{array} \right) ,
\end{equation}
where 
\begin{eqnarray}
  \epsilon_{\bf k} &=&-2t(\cos k_x + \cos k_y), \\
  m &=& \frac{1}{2} \langle a_{i\uparrow}^{\dagger}a_{i\uparrow} 
                           -a_{i\downarrow}^{\dagger}a_{i\downarrow}\rangle,
 \quad {\rm for}\,\, i\in A.
\end{eqnarray}
Matrix $\hat{J}({\bf k})$ is diagonalized by a unitary matrix
\begin{equation}
 \hat{U}({\bf k}) =
  \left( \begin{array}{cccc}
     \cos\theta_{\bf k} & -\sin\theta_{\bf k} & 0 & 0 \\
     \sin\theta_{\bf k} &  \cos\theta_{\bf k} & 0 & 0 \\
     0 & 0 & \cos\theta_{\bf k} & \sin\theta_{\bf k} \\
     0 & 0 & -\sin\theta_{\bf k} & \cos\theta_{\bf k}
         \end{array} \right),
\end{equation}
where the corresponding eigenvalues are $-E({\bf k})$, $E({\bf k})$, 
$E({\bf k})$, and $-E({\bf k})$ with 
$E({\bf k})=\sqrt{\epsilon_{\bf k}^2+(Um)^2}$, and $\theta_{\bf k}$ is
determined from
\begin{equation}
 \tan 2\theta_{\bf k} = -\frac{\epsilon_{\bf k}}{Um}.
\end{equation}
The sublattice magnetization $m$ is self-consistently determined by
\begin{equation}
 1 = U \frac{2}{N}\sum_{\bf k}\frac{1}{2E({\bf k})}.
\label{eq.gap2}
\end{equation}
For $U/t=4$ and $2$, $m$ is evaluated as
$0.345$ and $0.188$, respectively.

\subsection{Absorption coefficient}

\begin{figure}
\includegraphics[width=8.0cm]{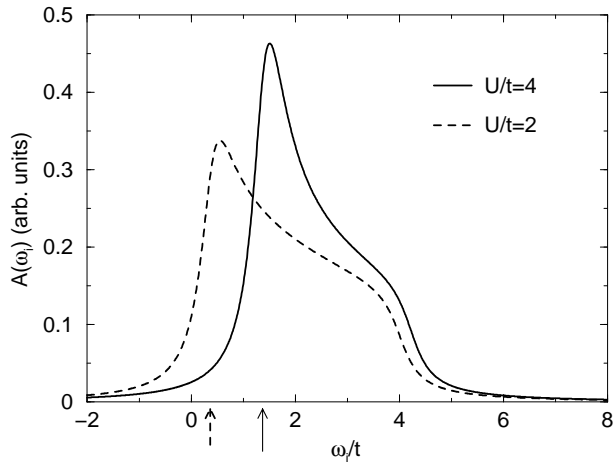}%
\caption{\label{fig.abs}
Absorption coefficient $A(\omega_i)$ as a function of x-ray energy,
for $U/t=4$ (solid line) and 2 (broken line). $\Gamma_c/t=0.3$.
The origin of energy is $\mu-\epsilon_{2p}(3/2)$, where $\mu$ is the chemical
potential lying at the middle of the energy gap in the single particle
energy band. The solid and broken arrows indicate 
the lowest bound of energy of unoccupied 
levels for $U/t=4$ and 2, respectively.
}
\end{figure}

X-ray could be absorbed by exciting  the $2p$ electron to unoccupied levels
at the $L$ edge. Neglecting the interaction between the excited electron and
the core hole left behind, we obtain the expression of the absorption
coefficient at the $L_3$ edge as
\begin{equation}
 A(\omega_i) \propto -\frac{2}{N}\sum_{\bf k}
   \frac{1}{\pi} \textrm{Im}[R(\omega_i,E(\textbf{k}))],
\end{equation}
where $\textrm{Im} X$ stands for the imaginary part of
the quantity $X$.
Figure \ref{fig.abs} shows the absorption coefficient $A(\omega_i)$ 
as a function of x-ray energy for $U/t=4$ and $2$. 
We set $\Gamma_c/t=0.3$.
The absorption peak is found at $\omega_i/t=1.51$ for $U/t=4$
and $\omega_i/t=0.55$ for $U/t=2$ where
$\omega_i$ is measured from $\mu-\epsilon_{2p}(3/2)$.

\subsection{RIXS spectra}

The transverse spin fluctuation caused by spin flips with changing
polarization of x-ray contains  the magnon as a bound state,
while the longitudinal spin fluctuation caused
without changing polarization could not give rise to the bound state. 
Since we are interested in the relation between the spectra arising 
from the bound state and the continuum states, we confine our study 
to the spin flip channel in the following.

Bearing a relation to cuprates in mind, we assume that the orbital 
describing the Hubbard model is $3d_{x^2-y^2}$.
In that situation, $M(\xi\xi';\alpha_i,\alpha_f)$ is the same as that given  
in the study of magnetic excitations in the L-edge RIXS from cuprates 
(Table II in Ref.~\onlinecite{Igarashi12-1}.).
With the spin quantization axis lying on the $xy$ plane and pointing to 
the direction at angle $\gamma$ with the $x$ axis, it is given by 
\begin{equation}
 M(-\sigma\sigma;y,x) = -M(-\sigma\sigma;x,y) = 
 \frac{i}{15}{\rm e}^{i\sigma\gamma}w^2,
\end{equation}  
where $w$ is a constant, and $\sigma=+1$ ($-1$) indicates the up 
(down) spin. When $\alpha_i$ and $\alpha_f$ are specified 
as $x$ or $y$, they are pointing to the $x$ or $y$, axes. 
Note that $M(\xi\xi';\alpha_i,\alpha_f)$ has no sublattice dependence.

In the spin flip channel, there exist two channels $c_1$ and $c_2$,
represented by the base states $(\xi\xi')=(A\uparrow A\downarrow)$,
$=(B\uparrow B\downarrow)$, called channel $c_1$,
and by the base states $(B\downarrow B\uparrow)$, $(A\downarrow A\uparrow)$,
called channel $c_2$.
In both channels, we have
\begin{eqnarray}
 \hat{M}(y,x) &=& -\hat{M}(x,y)=\frac{i}{15}w^2{\rm e}^{i\sigma\gamma}
               \left( \begin{array}{c}
                       1 \\
                       1 \end{array} \right) , \\
 \hat{\Gamma} &=& \left( \begin{array}{cc}
                          U & 0 \\
                          0 & U 
                         \end{array} \right) .
\end{eqnarray}
It is clear that both channels give rise to 
the same contributions to the RIXS intensities. 

We calculate the RIXS spectra by following the procedure in the preceding 
section. In channel $c_1$, for example, the zero-th order of the particle-hole 
Green's function is written as
\begin{eqnarray}
 \hat{F}({\bf q},\omega)&=&
\frac{2}{N}\sum_{\bf k}
   \frac{1}{\omega-E({\bf k+q})-E({\bf k})+i\delta} 
\nonumber \\
&\times& \left( \begin{array}{cc}
\sin^2\theta_{\bf k+q}\sin^2\theta_{\bf k} & -\frac{1}{4}
\sin 2\theta_{\bf k+q}\sin 2\theta_{\bf k} \\
-\frac{1}{4}\sin 2\theta_{\bf k+q}\sin 2\theta_{\bf k} &
\cos^2\theta_{\bf k+q}\cos^2\theta_{\bf k}
      \end{array} \right)
 \nonumber \\
&-& \frac{2}{N}\sum_{\bf k}\frac{1}
              {\omega+E({\bf k+q})+E({\bf k})-i\delta}
\nonumber \\
&\times&\left( \begin{array}{cc}
\cos^2\theta_{\bf k+q}\cos^2\theta_{\bf k} & -\frac{1}{4}
\sin 2\theta_{\bf k+q}\sin 2\theta_{\bf k} \\
-\frac{1}{4}\sin 2\theta_{\bf k+q}\sin 2\theta_{\bf k} &
\sin^2\theta_{\bf k+q}\sin^2\theta_{\bf k}
      \end{array} \right). \nonumber \\
\end{eqnarray}
The particle-hole Green's function is given by
$\hat{Y}^{T}({\bf q},\omega)=\hat{F}({\bf q},\omega)[\hat{I}+\hat{\Gamma}
\hat{F}({\bf q},\omega)]^{-1}$.

There exists a bound state below the energy continuum ($\omega< 2Um$).
The energy of the bound state should tend to $0$ with ${\bf q}\rightarrow 0$, 
since it is a Goldstone boson. This is proved as follows.
Let $\hat{F}({\bf q},\omega)$ be 
\begin{equation}
 \hat{F}({\bf q}\rightarrow 0,\omega\rightarrow 0) = \left(\begin{array}{cc}
                       f_1 & f_2 \\
                       f_2 & f_1 
                      \end{array} \right).
\end{equation}
Then the particle-hole Green's function is given by
\begin{equation}
 \hat{Y}^{T}({\bf q}\rightarrow 0,\omega\rightarrow 0) =
  \hat{S}\left(\begin{array}{cc}
               \frac{f_1+f_2}{1+U(f_1+f_2)} & 0 \\
               0 & \frac{f_1-f_2}{1+U(f_1-f_2)} 
               \end{array}\right)\hat{S}^{-1}, 
\end{equation}
where
\begin{equation}
  \hat{S} =\frac{1}{\sqrt{2}} \left(\begin{array}{cc}
                   1 & -1 \\
                   1 & 1\\
                  \end{array} \right) .
\end{equation}
Since 
\begin{equation}
 f_1-f_2 = -\frac{2}{N}\sum_{\bf k}
 \frac{(\sin^2\theta_{\bf k}+\cos^2\theta_{\bf k})^2}{2E_{\bf k}}
         =-\frac{2}{N}\sum_{\bf k}\frac{1}{2E_{\bf k}},
\end{equation}
we have $1+U(f_1-f_2)=0$ from Eq.~(\ref{eq.gap2}), indicating that 
a pole exist at ${\bf q}=0$ and $\omega=0$. In addition, we find from 
Eq.~(\ref{eq.wt.c}) that
\begin{equation}
  \hat{C}({\bf q}\rightarrow 0) \propto \left(\begin{array}{cc}
                                              \frac{1}{2} & -\frac{1}{2} \\
                                             -\frac{1}{2} & \frac{1}{2} 
                                              \end{array} \right) .
\end{equation}
In the fast collision approximation, the contribution to the RIXS intensity 
from the bound state vanishes with ${\bf q}\rightarrow 0$,
since $\hat{M}^{\dagger}\hat{C}\hat{M}=0$.
Beyond the fast collision approximation, however, it does not vanish 
due to the presence of $\hat{L}(\omega_i,q)$.

\begin{figure}
\includegraphics[width=8.0cm]{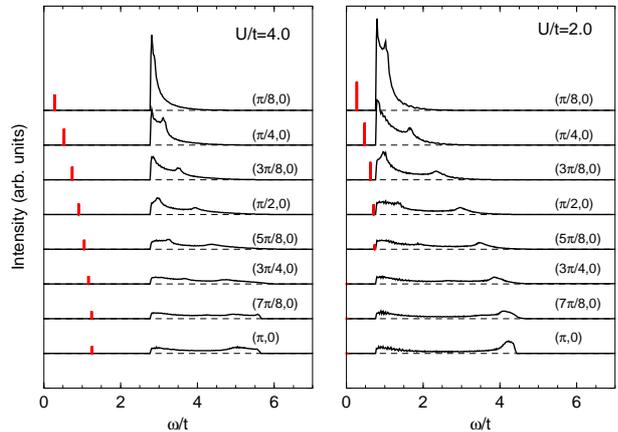}%
\caption{\label{fig.int}(Color Online)
RIXS spectra as a function of energy loss $\omega$ for the single-orbital
Hubbard model on the square lattice at half-filling 
along a symmetry line
${\bf q} \parallel (1,0)$. $\Gamma_c/t=0.3$.
The incident x-ray energy $\omega_i$ is set to give 
the maximum absorption coefficient; $\omega_i/t=1.51$ for $U/t=4$, and
$\omega_i/t=0.55$ for $U/t=2$ with $\omega_i$ measured from 
$\mu-\epsilon_{2p}(3/2)$.
Solid vertical line (red) indicates the magnon contribution.
}
\end{figure}

In the numerical calculation, we sum over wave vectors by dividing the first 
magnetic BZ into $128\times 128$ meshes.
Figure \ref{fig.int} shows the RIXS spectra as a function 
of $\omega$ along a symmetry line of ${\bf q} \parallel (1,0)$.
The energy of the incident x-ray $\omega_i$ is set to give 
the maximum absorption coefficient. We have the continuous spectra 
for $\omega> 2Um$, and below them we find $\delta$-function-like spikes arising from the magnon contribution.
For $U/t=4$, the magnon peak arising exists for all 
the $q$-values. Around ${\bf q}=0$, the velocity of magnon
$\omega/|{\bf q}|$ is estimated as $\sim 0.8$ in units of $t$,
which may be compared by $\sqrt{2}J$ with $J=4t^2/U$ for the localized
spin limit.
The intensity of the magnon peak decreases with increasing
value of $|{\bf q}|$; its value relative to the intensity of continuous 
states at $(\pi,0)$ is estimated as $0.175$ for $(\pi/8,0)$, $0.120$ for 
$(\pi/2,0)$, and $0.08$ for $(\pi,0)$. 
For $U/t=2$, the magnon mode approaches to the edge of the 
continuous spectra with increasing $|{\bf q}|$, and
it becomes difficult to judge numerically whether the magnon 
peak exists for $q_x > \pi/2$. 
The intensity of the magnon peak relative to the intensity of continuous 
states at $(\pi,0)$ is estimated as 0.27 for $(\pi/8,0)$ and 0.10 for 
$(\pi/2,0)$. 

\section{\label{sect.5}Concluding remarks}

We have developed a formalism of the $L$-edge RIXS spectra on a multi-orbital
tight-binding model. Without relying on the fast collision approximation,
we have derived the formulas useful to calculate the spectra,
by collecting up the ladder diagrams on the basis of the Keldysh scheme. 
Since the creation of an electron and that of a hole take place at different 
times, the formulas become complicated in comparison with the fast collision 
approximation. We think that, the present formulas, 
allowing to describe 
the spectra originated from both the bound state and the continuous states 
on the equal footing, work in the weak and intermediate coupling regime of 
itinerant electron systems, where the fast collision approximation loses 
ground with several orbitals involved with different energies. 
To demonstrate how the present formulas work, we have calculated the spectra
on a simple model, the single-orbital Hubbard model on 
the square lattice at half-filling. Now we know they work,
it may be interesting to apply the present formulas to analyze the spectra
of Sr$_2$IrO$_4$ on the itinerant electron model, although the analysis has 
already been carried out on the localized electron model
within the fast collision approximation.\cite{Ament11-1}

In the insulating phase of the strong coupling regime, 
the low-lying excitations may be described by the localized spin and/or orbital
model. In undoped cuptares, using the Heisenberg model, the RIXS spectra
arising from magnetic excitations have been calculated within the 
fast collision approximation\cite{Ament09,Haverkort10} 
and with using a ``projection" method to take account of the two-magnon 
excitations.\cite{Igarashi12-1,Igarashi12-2,Nagao12}
The present approximation scheme is unable to discuss such behavior in
the strong coupling.
To deal with the metallic phase with strong 
correlations, the present formalism for itinerant electron systems has to 
be extended by improving the HFA and the ladder approximation. 
In that extension, the single-particle Green's function is corrected by
the self-energy, and hence  the present formulas may be changed by 
noting that the Green's function within the HFA
\begin{equation}
 [\hat{G}({\bf k},\omega)]_{\xi_2\xi_1} =
 \sum_{j}\frac{U_{\xi_2 j}({\bf k})U_{\xi_1 j}^{*}({\bf k})}
              {\omega-E_j({\bf k})+i\delta{\rm sgn}[E_j({\bf k})]}
\end{equation}
is formally replaced by
\begin{equation}
 [\hat{G}({\bf k},\omega)]_{\xi_2\xi_1} 
  = \int{\rm d}\epsilon\frac{\rho_{\xi_2 \xi_1}({\bf k},\epsilon)}
    {\omega-\epsilon+i\delta{\rm sgn}[\epsilon]} .
\end{equation}
Such attempt may be related to the recent study on the RIXS spectra 
in the Falicov-Kimball model on the basis of the Keldysh scheme, 
\cite{Pakhira12}
where the dynamical mean field approximation is used.

\begin{acknowledgments}
This work was partially supported by a Grant-in-Aid for Scientific Research
from the Ministry of Education, Culture, Sports, Science, and Technology,
Japan. 
\end{acknowledgments}

\appendix
\section{\label{appen.A}Derivation of Eq.~(\ref{eq.L}) in the time 
representation}

\begin{figure}
\includegraphics[width=5.0cm]{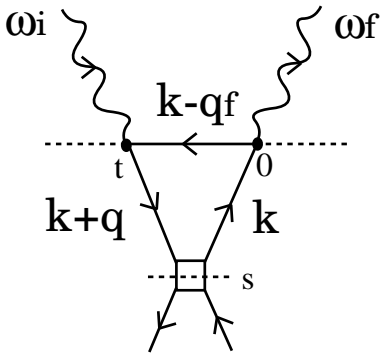}%
\caption{\label{fig.int.t}
Diagram for $\hat{L}(\omega_i;{\bf q},s)$ in the time representation.
The $s$ is the time at which the Coulomb interaction works.
}
\end{figure}

Let us consider a diagram in Fig.~\ref{fig.int.t} written 
in the time representation. 
The corresponding amplitude may be proportional to
\begin{eqnarray}
 \hat{L}(\omega_i;{\bf q},s)&\propto&\int_{-\infty}^{0} {\rm d}t
      iG_{2p}^{--}({\bf k-q}_f,t-0)iG^{--}({\bf k+q},s-t) \nonumber \\
&\times&
      iG^{--}({\bf k},0-s)\exp(-i\omega_i t),
\label{eq.int.t}
\end{eqnarray}
where the factors $U_{\xi_1 j}({\bf k+q})U^*_{\xi j}({\bf k+q})$ 
$U_{\xi' \ell}({\bf k})U^*_{\xi'_1 \ell}({\bf k})$, and the symbol of 
summing over ${\bf k}$ are omitted in this subsection.
The $\hat{L}(\omega_i;{\bf q},s)$ is the Fourier transform of
$\hat{L}(\omega_i,q=(\omega,{\bf q}))$.
Since $t<0$, we have
\begin{equation}
 iG_{2p}^{--}({\bf k-q}_f,t-0)=-\exp\{-i(\epsilon_{2p}(3/2)+i\Gamma_c)t\}.
\end{equation}

First we consider the time domain $s>0$. The Green's functions are given by
\begin{eqnarray}
 iG^{--}({\bf k},0-s)&=&-\sum_{\ell}n_{\ell}({\bf k})
\textrm{e}^{-iE_{\ell}({\bf k})(-s)},\\
 iG^{--}({\bf k+q},s-t)&=&\sum_{j}(1-n_{j}({\bf k+q}))
\textrm{e}^{-iE_j({\bf k+q}) (s-t)}.
\nonumber \\
\end{eqnarray}
Inserting these into Eq.~(\ref{eq.int.t}), we have 
\begin{eqnarray}
 \hat{L}(\omega_i;{\bf q},s) &\propto& 
 -\frac{1}{i} R(\omega_i,E_j({\bf k+q}))
  (1-n_j({\bf k+q}))n_{\ell}({\bf k})
\nonumber \\
&\times& \exp\{-i(E_j({\bf k+q})-E_{\ell}({\bf k}))s\}.
\end{eqnarray}
The Fourier transform of this function corresponds to the former part of 
the second term of Eq.~(\ref{eq.L}).

\begin{widetext}
Next we consider the time domain $s<0$. The Green's functions are given by
\begin{eqnarray}
iG^{--}({\bf k},0-s)&=&\sum_{\ell}(1-n_{\ell}({\bf k}))
  \textrm{e}^{-iE_{\ell}({\bf k})(-s)},\\
iG^{--}({\bf k+q},s-t)&=&
 \left\{\begin{array}{ll}
       -\sum_{j} n_{j}({\bf k+q})
        \textrm{e}^{-iE_j({\bf k+q})(s-t)}, &  s<t,\\
       \sum_{j}(1-n_{j}({\bf k+q}))
        \textrm{e}^{-iE_j({\bf k+q})(s-t)}, & t<s<0.
        \end{array}
 \right.
\end{eqnarray}
Inserting these into Eq.~(\ref{eq.int.t}), we have 
\begin{eqnarray}
 \hat{L}(\omega_i;{\bf q},s) &\propto& 
  -\frac{1}{i} R(\omega_i,E_j({\bf k+q}))
 \Bigl\{n_j({\bf k+q}))(1-n_{\ell}({\bf k}))
 \exp\{-i(E_j({\bf k+q})-E_{\ell}({\bf k}))s\}\nonumber\\
       &-&(1-n_{\ell}({\bf k}))
       \exp\{-i(\omega_i+\epsilon_{2p}(3/2)+i\Gamma_c
             -E_{\ell}({\bf k}))s\}
       \Bigr\}.
\end{eqnarray}
The Fourier transform of the first term corresponds to the first term of 
Eq.~(\ref{eq.L}), and that of the second term corresponds to the latter part 
of the second term of Eq.~(\ref{eq.L}).
\end{widetext}

\section{\label{appen.B}Derivation of Eq.~(\ref{eq.fdt1})}

Let $\hat{A}_1$ and $\hat{A}_2$ be $\rho_{{\bf q}\xi_1\xi'_{1}}$ and
$\rho_{{\bf q}\xi\xi'}$, respectively. Then we have
\begin{eqnarray}
 \left[\hat{Y}^{T}(\omega)\right]_{1,2} &\equiv& -i\int_{-\infty}^{\infty}
  \langle T(A_{1}^{\dagger}(t)A_{2}(0))\rangle {\rm e}^{i\omega t}{\rm d}t 
 \nonumber\\
  &=& \sum_{n} 
      \frac{\langle 0|A_1^{\dagger}|n\rangle\langle n|A_2|0\rangle}
           {\omega - E_n + E_0 + i\delta} \nonumber\\
  &-& \sum_{n} 
      \frac{\langle 0|A_2|n\rangle\langle n|A_1^{\dagger}|0\rangle}
           {\omega + E_n - E_0 - i\delta}, \\
 \left[\hat{Y}^{T}(\omega)\right]^{*}_{2,1} 
  &=& \sum_{n} 
      \frac{\langle 0|A_1^{\dagger}|n\rangle\langle n|A_2|0\rangle}
           {\omega - E_n + E_0 - i\delta} \nonumber\\
  &-& \sum_{n} 
      \frac{\langle 0|A_2|n\rangle\langle n|A_1^{\dagger}|0\rangle}
           {\omega + E_n - E_0 + i\delta}, 
\end{eqnarray}
where $|0\rangle$ and $|n\rangle$ denote the ground state with energy 
$E_0$ and the excited state with energy $E_n$, respectively.
Hence we have
\begin{eqnarray}
 & & \left[\hat{Y}^{T}(\omega)\right]^{*}_{2,1}-
 \left[\hat{Y}^{T}(\omega)\right]_{1,2} \nonumber\\
  &=& 2\pi i\sum_{n} \langle 0|A_1^{\dagger}|n\rangle\langle n|A_2|0\rangle
               \delta(\omega-E_n+E_0) \nonumber\\
  &+& 2\pi i\sum_{n} \langle 0|A_2|n\rangle\langle n|A_1^{\dagger}|0\rangle
               \delta(\omega+E_n-E_0) .
\end{eqnarray} 
Comparing this with the expression 
\begin{eqnarray}
 \left[\hat{Y}^{+-}(\omega)\right]_{1,2} 
&\equiv& \int_{-\infty}^{\infty}
  \langle A_{1}^{\dagger}(t)A_{2}(0)\rangle {\rm e}^{i\omega t}{\rm d}t 
 \nonumber\\
  &=& 2\pi\sum_{n} \langle 0|A_1^{\dagger}|n\rangle\langle n|A_2|0\rangle
           \delta(\omega - E_n + E_0),
\nonumber \\
\end{eqnarray}
we have $[\hat{Y}^{+-}(\omega)]_{1,2}
=-i\{[\hat{Y}^{T}(\omega)]^{*}_{2,1}
-[\hat{Y}^{T}(\omega)]_{1,2}\}$ for $\omega>0$.

\bibliographystyle{apsrev} 
\bibliography{paper}

\end{document}